\documentclass[reprint,prx,twocolumn,showpacs,superscriptaddress,aps,longbibliography]{revtex4-2}

\usepackage[colorlinks=true,citecolor=blue,linkcolor=blue]{hyperref} 
\usepackage{graphicx}
\usepackage{bm}
\usepackage{amsmath}
\usepackage{amssymb}

\DeclareGraphicsExtensions{.png,.jpg,.eps}
\usepackage{xcolor}

\newcommand{\beq}{\begin{equation}}
\newcommand{\eeq}{\end{equation}}

\usepackage{mleftright} 

\newcommand{\secref}[1]{\mbox{Section~\ref{#1}}}

\renewcommand{\eqref}[1]{\mbox{Eq.~(\ref{#1})}}

\newcommand{\figpanel}[2]{Fig.~\hyperref[#1]{\ref*{#1}(#2)}}
\newcommand{\figpanels}[3]{Fig.~\hyperref[#1]{\ref*{#1}(#2)-(#3)}}
\newcommand{\figpanelNoPrefix}[2]{\hyperref[#1]{\ref*{#1}(#2)}}

\newcommand{\ket}[1]{|#1 \rangle}
\newcommand{\bra}[1]{\langle #1|}

\usepackage{siunitx}

\begin{document}

\author{Guangze Chen}
\email{guangze@chalmers.se}
\affiliation{Department of Microtechnology and Nanoscience, Chalmers University of Technology, 41296 Gothenburg, Sweden}

\author{Anton Frisk Kockum}
\email{anton.frisk.kockum@chalmers.se}
\affiliation{Department of Microtechnology and Nanoscience, Chalmers University of Technology, 41296 Gothenburg, Sweden}


\title{Efficient three-qubit gates with giant atoms}

\begin{abstract}

Three-qubit gates are highly beneficial operations in quantum computing, enabling compact implementations of quantum algorithms and efficient generation of multipartite entangled states. However, realizing such gates with high fidelity remains challenging due to crosstalk, complex control requirements, and the overhead of parametric or tunable couplers. In this work, we propose and analyze the implementation of fast, high-fidelity three-qubit gates using giant atoms---artificial atoms coupled to a waveguide at multiple spatially separated points. By leveraging interference effects intrinsic to the giant-atom architecture, we demonstrate that native three-qubit gates, such as the controlled-CZ-SWAP (CCZS) and the dual-iSWAP (DIV), can be realized through simple frequency tuning, without the need for complex pulse shaping or additional hardware. We evaluate gate performance under realistic decoherence and show that fidelities exceeding $\qty{99.5}{\percent}$ are achievable with current experimental parameters in superconducting circuits. As an application, we present a scalable protocol for preparing three- and five-qubit GHZ states using minimal gate depth, achieving high state fidelity within sub-\qty{300}{\nano\second} timescales. Our results position giant-atom systems as a promising platform for entangled-state preparation and low-depth quantum circuit design in near-term quantum computers and quantum simulators.

\end{abstract}

\date{\today}

\maketitle


\section{Introduction}

Quantum computers~\cite{nielsen_chuang_2010, Feynman1982} hold the promise of solving complex problems that are practically intractable for classical machines, with potential applications across a variety of disciplines~\cite{Cao2019, McArdle2020, Georgescu2014, Wendin2017, Preskill2018quantumcomputingin, Cerezo2021, Dalzell2023}. However, state-of-the-art and near-future devices fall into the category of noisy intermediate-scale quantum (NISQ) computers~\cite{Preskill2018quantumcomputingin}, where decoherence and control errors limit circuit depth and restrict the reliable execution of complex quantum algorithms. Many quantum algorithms, particularly those targeting quantum simulation~\cite{Georgescu2014, Altman2021, Fauseweh2024}, require the generation of large-scale entanglement or the implementation of effective many-body interactions. When such operations are decomposed into sequences of two-qubit gates, the resulting quantum circuit depth increases substantially, thereby amplifying the effects of noise and reducing fidelity~\cite{Abad2022, Abad2025}. 

In this context, native three-qubit and multi-qubit gates offer a compelling alternative: they can reduce circuit depth~\cite{Leymann2020, kalloor2024quantumhardwarerooflineevaluating, Ge2024, tasler2025optimizingsuperconductingthreequbitgates}, operate with similar or even shorter durations than two-qubit gates~\cite{Gu2021, Warren2023}, and thereby ultimately yield higher fidelities---an important advantage in the NISQ era. These capabilities are particularly useful for preparing highly entangled states such as Greenberger--Horne--Zeilinger (GHZ)~\cite{Greenberger1989, PhysRevLett.82.1345} and Dicke~\cite{PhysRev.93.99} states, which are key resources in quantum information processing and quantum simulation~\cite{Nielsen2000, Chitambar2019, Zhao2025}.

Despite their attractive features, implementing high-fidelity three-qubit gates remains a significant challenge. Most experimental realizations decompose them into single- and two-qubit gates~\cite{PhysRevX.14.041030, Sun2024}, as native implementations demand greater hardware complexity and fine-grained control. In superconducting qubit platforms~\cite{Gu2017, Blais2021}, existing schemes rely on parametric or flux-tunable couplers~\cite{Warren2023, PhysRevApplied.21.044035, lvb9-pfr3}, intricate drive protocols~\cite{PhysRevApplied.19.044001, Su2023, Huai2024, Li2024, PhysRevApplied.21.034018}, or hardware-specific architectures tailored to particular gates~\cite{Roy2025}. These approaches often require extensive calibration, are sensitive to timing and detuning errors, and suffer from frequency crowding and crosstalk, which limit scalability. Trapped-ion~\cite{Strohm2024} implementations, while capable of high-fidelity operations, are constrained by motional heating and control complexity in large systems~\cite{schwerdt2025opticaltweezercontrolledentanglementgates, Ringbauer2022, Hrmo2023}. In neutral-atom platforms~\cite{Wintersperger2023}, Rydberg blockade-based three-qubit gates face tight requirements on spatial alignment and uniform coupling strengths~\cite{Omran2019}.

To address these challenges, we propose a new approach to three-qubit gates based on giant artificial atoms~\cite{Kockum2021, Gustafsson2014, Kockum2014, Kockum2018, Kannan2020}---superconducting qubits or other quantum emitters that couple to their environment (e.g., a waveguide) at multiple discrete points spaced by wavelengths distances. Such systems have been studied for other purposes both theoretically~\cite{Kockum2014, Guo2017, Kockum2018, Gonzalez-Tudela2019, Guo2020, Guimond2020, Ask2020, Cilluffo2020, Wang2021, Du2021, Soro2022, Wang2022, Du2022, Du2022a, Terradas-Brianso2022, Soro2023, Du2023, Ingelsten2024, Leonforte2024, Wang2024, Roccati2024, Gong2024, Du2025, Du2025a} and experimentally~\cite{Gustafsson2014, Manenti2017, Satzinger2018, Bienfait2019, Andersson2019, Kannan2020, Bienfait2020, Andersson2020, Vadiraj2021, Wang2022a, Joshi2023, Hu2024, Almanakly2025} since their inception a decade ago. Unlike conventional ``small'' atoms that interact locally, giant atoms exhibit nonlocal coupling, leading to interference effects that allow frequency-dependent control over both relaxation rates~\cite{Kockum2014, Kannan2020, Vadiraj2021, Wang2022a} and qubit-qubit interaction strengths~\cite{Kockum2018, Kannan2020, Wang2022a}.  This frequency control allows for versatile gate operations without the need for parametric modulation~\cite{Kannan2020, Chen2025, chen2025scalablequantumsimulatorextended}. Crucially, it makes giant atoms a natural platform for simulating open quantum systems~\cite{Chen2025, chen2025scalablequantumsimulatorextended}, where diverse gate operations—including controlled dissipation and nonlocal interactions—must be realized. Incorporating native three-qubit gates in giant atoms would substantially enhance their utility. Yet, such gates have not previously been realized in this platform.

In this article, we demonstrate how two important three-qubit gates---the controlled-CZ-SWAP (CCZS) and dual-iSWAP (DIV) gates~\cite{Gu2021, Warren2023}---can be implemented natively in giant-atom systems. We analyze the gate fidelities in these implementations under realistic decoherence and coupling strengths, finding that fidelities exceeding $\qty{99.5}{\percent}$ are attainable with current technology. As an application of these gates, we present protocols for the fast and scalable preparation of both three- and five-qubit GHZ states. These results establish giant atoms as a powerful platform for entangled-state preparation and low-overhead circuit design, particularly relevant to the simulation of open-system dynamics.


\section{Three-qubit gates with giant atoms} \label{sec2}

In this section, we describe the setup with giant atoms that we propose to use for both three-qubit gates we consider in this article. We then describe the implementation and numerical simulation of first the CCZS and then the DIV gate, showing that both gates can be implemented with high fidelity for realistic parameter choices in superconducting qubits.


\subsection{Setup}

\begin{figure}
\center
\includegraphics[width=\linewidth]{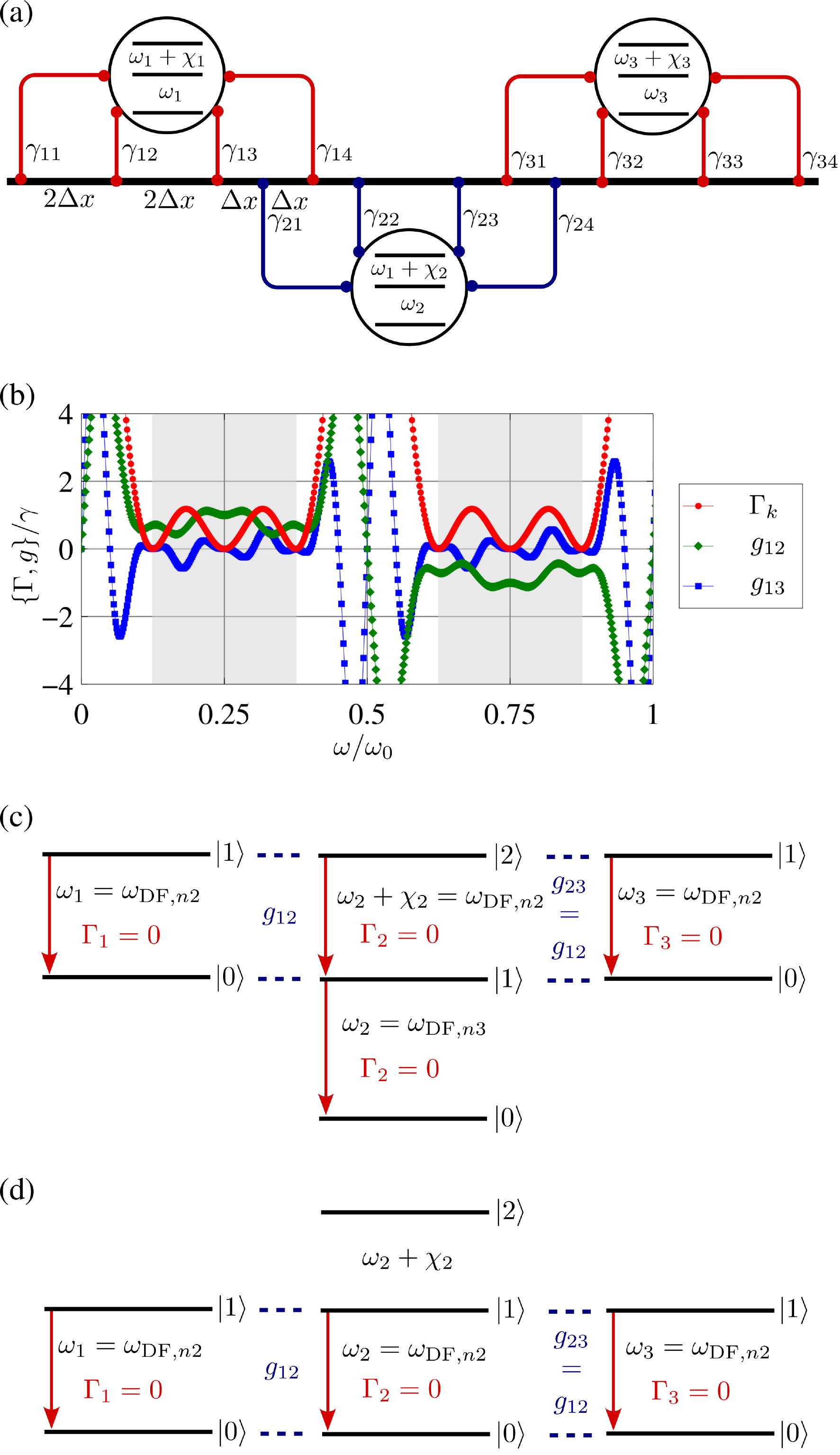}
\caption{Three-giant-atom setup for implementing CCZS and DIV gates.
(a) Schematic of the system. Three giant atoms with transition frequencies $\omega_1$, $\omega_2$, and $\omega_3$, and anharmonicities $\chi_1$, $\chi_2$, and $\chi_3$, are coupled to a waveguide (black line) at multiple spatially separated points with coupling strengths $\gamma$. The coupling points are organized in a braided fashion.
(b) Frequency dependence of the individual decay rates $\Gamma_\text{ind}$ and inter-atomic coupling strengths $g_{12}$ and $g_{13}$, showing selective activation of desired interactions. Due to the symmetry of the setup, $g_{12} = g_{23}$.
(c,d) Frequency configurations used to perform (c) a CCZS gate and (d) a DIV gate, exploiting interference effects at decoherence-free points.}
\label{fig1}
\end{figure}

We consider a setup of three giant artificial atoms that enables the implementation of three-qubit gates through simple frequency tuning. The giant atoms we consider are ladder-type ($\Xi$-type) three-level systems with negative anharmonicity $\chi_{k} = \omega^{(2)}_{k} - \omega^{(1)}_{k} <0$, where $\omega^{(\alpha)}_{k}$ is the transition frequency between states $\ket{\alpha}_k$ and $\ket{\alpha-1}_k$ in atom $k$. We denote $\omega^{(1)}_{k} \equiv \omega_k$ for convenience. This level structure is typical for superconducting transmon qubits~\cite{Koch2007}, the most common experimental platform for giant atoms. The giant atoms are coupled to a common waveguide at multiple spatially separated points, with coupling rate $\gamma_{kn}$ at the $n$th coupling point of atom $k$ [\figpanel{fig1}{a}]. In the Markovian limit, and assuming a linear waveguide dispersion with propagation velocity $v$, the dynamics of the atoms are governed by the master equation~\cite{Kockum2018} ($\hbar = 1$ throughout this manuscript)
\begin{widetext}
\begin{align} \label{eq_setup}
\partial_t \rho &= -i \mleft[ H , \rho \mright] + \sum_k\sum^2_{\alpha=1}\Gamma_k\mleft(\omega^{(\alpha)}_k\mright) \mathcal{D}\mleft[\sigma_k^{(\alpha),-}\mright] \rho \nonumber \\
& \quad + \sum_{j\neq k}\sum_{\alpha,\beta}\Gamma_{jk}\mleft(\omega^{(\alpha)}_j,\omega^{(\beta)}_k\mright) \mleft[ \mleft( \sigma_j^{(\alpha),-} \rho \sigma_k^{(\beta),+} - \frac{1}{2} \mleft\{ \sigma_j^{(\alpha),+} \sigma_k^{(\beta),-} , \rho \mright\} \mright) + \text{H.c.} \mright],
\end{align}
\end{widetext}
where the coherent dynamics is generated by the Hamiltonian
\begin{align} \label{eq_setup_H}
H= & \sum_k \mleft( \omega^{(1)}_k \ket{1}_k\bra{1}_k +  \mleft( \omega^{(1)}_k + \omega^{(2)}_k\mright) \ket{2}_k\bra{2}_k \mright) \nonumber \\
& + \sum_{j,k}\sum_{\alpha,\beta}g_{jk}\mleft(\omega^{(\alpha)}_j,\omega^{(\beta)}_k\mright)\mleft( \sigma_j^{(\alpha),-}\sigma_k^{(\beta),+} + \text{H.c.}\mright).
\end{align}
Here, 
\begin{align}
\mathcal{D}\mleft[\sigma_k^{(\alpha),-}\mright] \rho = \sigma_k^{(\alpha),-} \rho \sigma_k^{(\alpha),+} - \frac{1}{2}\mleft\{ \sigma_k^{(\alpha),+} \sigma_k^{(\alpha),-}, \rho\mright\}
\end{align}
is the Lindblad dissipator, and the lowering and raising operators are defined as
\beq
\sigma_k^{(\alpha),-}\ket{\alpha}_k=\sqrt{\alpha}\ket{\alpha-1}_k, \quad \sigma_k^{(\alpha),+}=\mleft[\sigma_k^{(\alpha),-}\mright]^\dag.
\eeq
The decay rates of the atoms and their interaction strengths are given by~\cite{Kockum2018}
\beq \label{eq_rates}
\begin{aligned}
\Gamma_{k}(\omega)=\sum_{n=1}^{N_k}\sum_{m=1}^{N_k}\sqrt{\gamma_{kn}\gamma_{km}}\cos\phi_{kn,km}(\omega),\\
g_{jk}(\omega,\omega)=\sum_{n=1}^{N_j}\sum_{m=1}^{N_k}\frac{\sqrt{\gamma_{jn}\gamma_{km}}}{2}\sin\phi_{jn,km}(\omega),\\
\Gamma_{jk}(\omega,\omega)=\sum_{n=1}^{N_j}\sum_{m=1}^{N_k}\sqrt{\gamma_{jn}\gamma_{km}}\cos\phi_{jn,km}(\omega),
\end{aligned}
\eeq
where $\Gamma_{k}(\omega)$ are individual decay rates for atom $k$, $g_{jk}(\omega)$ is the coherent interaction strength between atoms $j$ and $k$, and $\Gamma_{jk}(\omega)$ denotes collective decay rate. Here, $N_k$ is the number of coupling points of atom $k$ and $\phi_{jn,km}(\omega) = \omega\Delta x_{jn,km}/v$ is the phase difference accumulated between the $n$th coupling point of atom $j$ and the $m$th coupling point of atom $k$ with $\Delta x_{jn,km}$ being their distance.

The multiple coupling points of giant atoms result in interference effects that affect the decay and interaction rates in \eqref{eq_rates}. Remarkably, when the coupling points are arranged in a ``braided'' fashion, i.e., some coupling point(s) of each atom are placed between some of the coupling points of another atom [\figpanel{fig1}{a}], the interference effects can give rise to decoherence-free interactions~\cite{Kockum2018}. In this regime, the atomic decays $\Gamma_k$ and $\Gamma_{jk}$ vanish, while the coherent interaction $g_{jk}$ remains non-zero for specific pairs of $j,k$. To illustrate this, we consider identical coupling strengths $\gamma_{kn}=\gamma$. As shown in \figpanel{fig1}{b}, this configuration supports a sequence of decoherence-free frequencies $\omega_{\text{DF},nm}=(n+m/8)\omega_0$ $(n\in\mathcal{N}, m=1,2,3,5,6,7)$, where $\omega_0 = 2 \pi v / \Delta x$ and $\Delta x$ is the distance between coupling points illustrated in \figpanel{fig1}{a}. At these decoherence-free frequencies, the interaction $g_{12}$ ($= g_{23}$ due to the symmetry of the setup) is non-zero, while the unwanted interaction $g_{13}$ is zero. This arrangement of coupling points enables precise, high-fidelity, and selective multi-qubit gates. The individual decay remains low over frequency intervals around these points, forming robust operational windows [shaded regions in \figpanel{fig1}{b}] which can be widened by adding more coupling points. 

This setup, previously used for high-fidelity two-qubit iSWAP and CZ gates~\cite{chen2025scalablequantumsimulatorextended}, also supports direct implementation of three-qubit gates. In the following sections, we demonstrate how it enables native realization of CCZS and DIV gates~\cite{Gu2021}, and use them to efficiently generate multi-qubit entangled states.

For the case of the CCZS gate (but not for the DIV gate), we will need to use the second excited state of the middle qubit. To suppress unwanted transitions during gate operations in this case, we set $\chi_1\neq\chi_2=-\chi\neq\chi_3$, where $\chi=\omega_0/8$. Note that $\omega_0$ is not the transition frequency of any qubit, but can be chosen using the distance between coupling points such that this anharmonicity matches that of transmon qubits (by using $n > 1$); if a larger anharmonicity is needed, that could be achieved by using flux qubits instead~\cite{Yan2016}.


\subsection{CCZS gate with giant atoms}
\label{sec:CCZS}

The CCZS gate~\cite{Gu2021} is a native 3-qubit gate where one qubit controls a simultaneous CZ and SWAP (CZS) gate on the other two qubits. It can be achieved by having the following transitions activated in the 3-qubit system: $\ket{120}\leftrightarrow\ket{111}\leftrightarrow\ket{021}$ and $\ket{110}\leftrightarrow\ket{020}\leftrightarrow\ket{011}$, with the second qubit acting as the control. Note that having just the transitions on the left between qubits 1 and 2, or just the transitions on the right between qubits 2 and 3, is one way that is used to implement two-qubit CZ gates between those qubits. 

When all the transition rates are the same ($g$), the gate duration is $t_{\rm CCZS}=\pi/(\sqrt{2}g)$ (the gate duration for a CZ gate between two qubits in the same setup would be $t_{\rm CZ} = \pi / g$, i.e., a factor $\sqrt{2}$ slower), and the CCZS gate implements the unitary transformation
\beq
U_{\rm CCZS}=\ket{0}_2\bra{0}_2\otimes\mathbf{1}_1\otimes\mathbf{1}_3+\ket{1}_2\bra{1}_2\otimes \mleft(U_{\rm CZS}\mright)_{13} ,
\eeq
where
\beq
U_{\rm CZS}=\mleft(\begin{array}{cccc}
    1 \\ &&-1 \\ &-1\\ &&&-1 
\end{array}\mright)
\eeq
is a simultaneous CZ and SWAP gate.

While this coupling scheme can be engineered using parametric couplers~\cite{Warren2023}, such implementations require additional hardware and complex modulation schemes. In contrast, our giant-atom setup enables the same interaction without auxiliary circuitry, by tuning the qubit frequencies to $\omega_1=\omega_3=\omega_{\text{DF},n2}$ and $\omega_2=\omega_{\text{DF},n3}$; see \figpanel{fig1}{c}. This choice selectively activates couplings $g_{12}=g_{23}=g=\gamma$, while leaving the unwanted $g_{13}$ zero, thereby enabling the desired transitions purely through frequency control.

\begin{figure}
\center
\includegraphics[width=\linewidth]{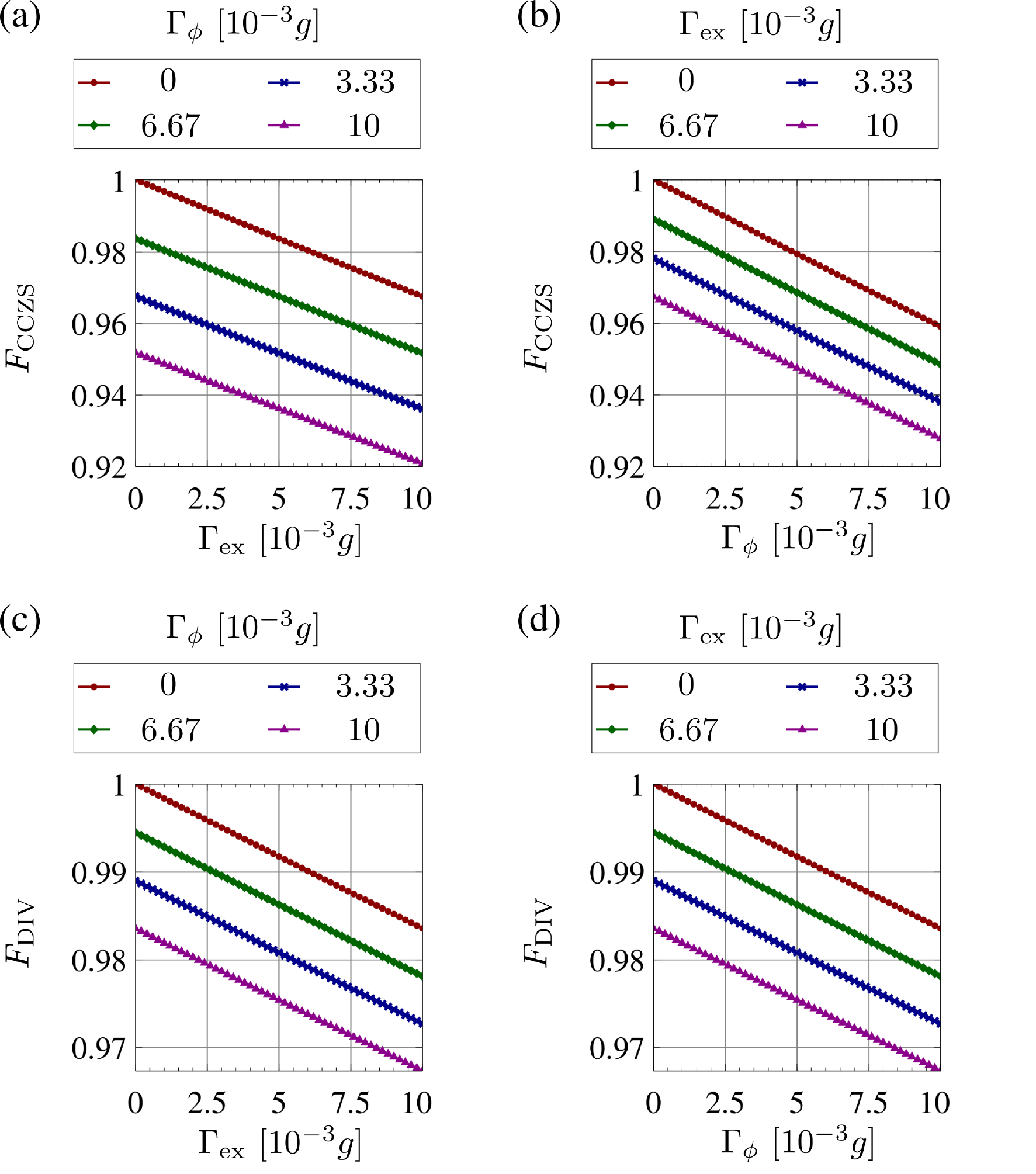}
\caption{Average process fidelity of three-qubit gates implemented with the giant-atom setup shown in \figpanel{fig1}{a}. 
(a) Fidelity of the CCZS gate as a function of qubit decay rate $\Gamma_\text{ex}$ and (b) dephasing rate $\Gamma_\phi$. 
(c,d) Fidelity of the DIV gate under the same noise conditions. In all cases, the qubit–qubit coupling strength is $g=\gamma$, set by the waveguide-mediated interaction.}
\label{fig2}
\end{figure}

We now analyze the fidelity of the giant-atom CCZS gate in realistic cases, when the qubits are subject to decay and dephasing. To be concrete, we consider the case of transmon qubits. The Lindblad jump operators for qubit decay and dephasing are~\cite{chen2025scalablequantumsimulatorextended}
\begin{align}
L_{-,k} &=\sqrt{\Gamma_{k} + \Gamma_{\text{ex},k}} \mleft( \sigma^{(1),-}_k + \sigma^{(2),-}_k \mright), \\
L_{\phi,k} &=\sqrt{2\Gamma_{\phi,k}} \mleft( \ket{1}_k\bra{1}_k + 2 \ket{2}_k\bra{2}_k \mright) ,
\end{align}
respectively, where $\Gamma_{\text{ex},k}$ is the extra decay rate of qubit $k$ to environments other than the waveguide, and $\Gamma_{\phi,k}$ is the qubit's dephasing rate. For simplicity, we further assume that the qubits have the same decay and dephasing rates: $\Gamma_{\text{ex},k}=\Gamma_{\rm ex}$ and $\Gamma_{\phi,k}=\Gamma_\phi$. The dynamics of the qubits is governed by the master equation
\begin{align} \label{eq_CCZS_dynamics}
\partial_t \rho &= -i \mleft[ g\mleft(\sigma^{(1),-}_1\sigma^{(2),+}_2 + \sigma^{(1),-}_3\sigma^{(2),+}_2 + \text{H.c.} \mright) , \rho \mright] \nonumber \\ 
& + \sum_k \mleft(\mathcal{D}\mleft[L_{-,k}\mright] + \mathcal{D}\mleft[L_{\phi,k}\mright]\mright) \rho.
\end{align}

We quantify performance for the three-qubit gate using the average process fidelity~\cite{Gilchrist2005}:
\begin{equation}
F_\text{CCZS}=\mleft[\text{tr}\mleft(\sqrt{\sqrt{\Phi}\Phi_0\sqrt{\Phi}}\mright)\mright]^2 ,
\end{equation}
where $\Phi$ and $\Phi_0$ are the Choi matrices~\cite{Choi1975} of the implemented and ideal gates, respectively. The Choi matrix $\Phi$ of a process $\mathcal{E}$ is defined as
\beq \label{eq_Choi}
\Phi = \sum_n \ket{n}\bra{n} \otimes \mathcal{E}(\ket{n}\bra{n}),
\eeq
where $n$ runs over the basis states of the computational subspace. In our case, this subspace is spanned by $\ket{0}$ and $\ket{1}$ states of the qubits, and has a dimension of $2^3=8$. The Choi matrix of the implemented gate is obtained by numerically evolving \eqref{eq_CCZS_dynamics} for a gate duration $t_{\rm CCZS}=\pi/(\sqrt{2}g)$, which corresponds to the ideal CCZS operation in the absence of dissipation.

As shown in \figpanels{fig2}{a}{b}, the fidelity scales linearly with decoherence to a good approximation: $F_\text{CCZS}\approx1-3.26\Gamma_\text{ex}/g-4.09\Gamma_\phi/g$, corresponding to an average gate fidelity $F_{\text{ave, CCZS}}\approx1-2.90\Gamma_\text{ex}/g-3.64\Gamma_\phi/g$~\footnote{The average gate fidelity and the process fidelity are related as~\cite{Gilchrist2005}: $F_\text{ave}=1-dF/(d+1)$ where $F_\text{ave}$ and $F$ are the average gate and process fidelities, and $d$ is the dimension of the computational space. For the three-qubit CCZS gate, we have $d=8$.}. For realistic parameteres of $\gamma/ (2\pi) = \qty{4}{\mega\hertz}$, $\Gamma_{\rm ex} = \qty{0.01}{\mega\hertz}\approx 0.40 \cdot 10^{-3}g$ and $\Gamma_\phi = \qty{0.02}{\mega\hertz} \approx 0.80 \cdot 10^{-3}g$~\cite{annurev_qubits, Place2021, Somoroff2023, Kim2023, Biznarova2023, Kono2024, Bal2024}, the average gate fidelity is $\qty{99.59}{\percent}$. Higher fidelities can be achieved by increasing $\gamma$, which proportionally reduces the gate time and thus the impact of decoherence.


\subsection{DIV gate with giant atoms}

The DIV gate is another three-qubit operation where simultaneous activation of transitions used individually for two-qubit gates results in a more powerful total operation~\cite{Gu2021}. This gate works by activating the iSWAP transitions of qubit pairs (1, 2) and (2, 3), i.e., $\ket{010}\leftrightarrow\ket{100}\leftrightarrow\ket{001}$ and $\ket{101}\leftrightarrow\ket{010}\leftrightarrow\ket{110}$. 

Because the total excitation number is conserved, the full unitary can be written as $U_{DIV}=\mathbf{1}_{(0)}\oplus U_{(1)} \oplus U_{(2)} \oplus \mathbf{1}_{(3)}$ where the subscripts label the $n$-excitation subspaces. The 0- and 3-excitation sectors remain unchanged, while the 1- and 2-excitation sectors evolve under the same three-level unitary $U$. When all the transitions have the same rate $g$, the gate time for the DIV gate is $t_{\rm DIV}=\pi/(2\sqrt{2}g)$ and the unitary operation is
\beq
U=\mleft(\begin{array}{ccc}
    1/2 & -i/\sqrt{2} & -1/2 \\ -i/\sqrt{2}&0&-i/\sqrt{2} \\ -1/2&-i/\sqrt{2}&1/2
\end{array}\mright).
\eeq

The DIV coupling pattern is realized in our architecture by tuning all three qubits to the same decoherence-free frequency, $\omega_1=\omega_2=\omega_3=\omega_{\text{DF},n2}$; see \figpanel{fig1}{d}. This frequency setting activates couplings $g_{12}=g_{23}=g=\gamma$, while suppressing unwanted $g_{13}$ and individual decays, enabling clean three-body dynamics without additional hardware.

We evaluate the performance of the DIV gate under the same decoherence model as we used for CCZS in \secref{sec:CCZS}. As can be seen in \figpanels{fig2}{c}{d}, the shorter gate time of the DIV gate results in higher process fidelity than for the CCZS gate: $F_\text{DIV}\approx1-1.64\Gamma_\text{ex}/g-1.64\Gamma_\phi/g$. Correspondingly, the average gate fidelity is $F_{\text{ave, DIV}}\approx1-1.46\Gamma_\text{ex}/g-1.46\Gamma_\phi/g$. For representative parameters of $\gamma/ (2\pi) = \qty{4}{\mega\hertz}$, $\Gamma_{\rm ex} = \qty{0.01}{\mega\hertz}$ and $\Gamma_\phi = \qty{0.02}{\mega\hertz}$, the average gate fidelity is $\qty{99.83}{\percent}$. 


\section{Preparation of GHZ states with giant atoms} \label{sec3}

As an application of the three-qubit gates enabled by our giant-atom platform, we demonstrate the fast and high-fidelity preparation of GHZ states. Prior work~\cite{Gu2021} has shown that the CCZS gate enables faster entanglement generation compared to sequences of two-qubit gates, making it particularly well suited for constructing GHZ states with reduced circuit depth.


\begin{figure}
\center
\includegraphics[width=\linewidth]{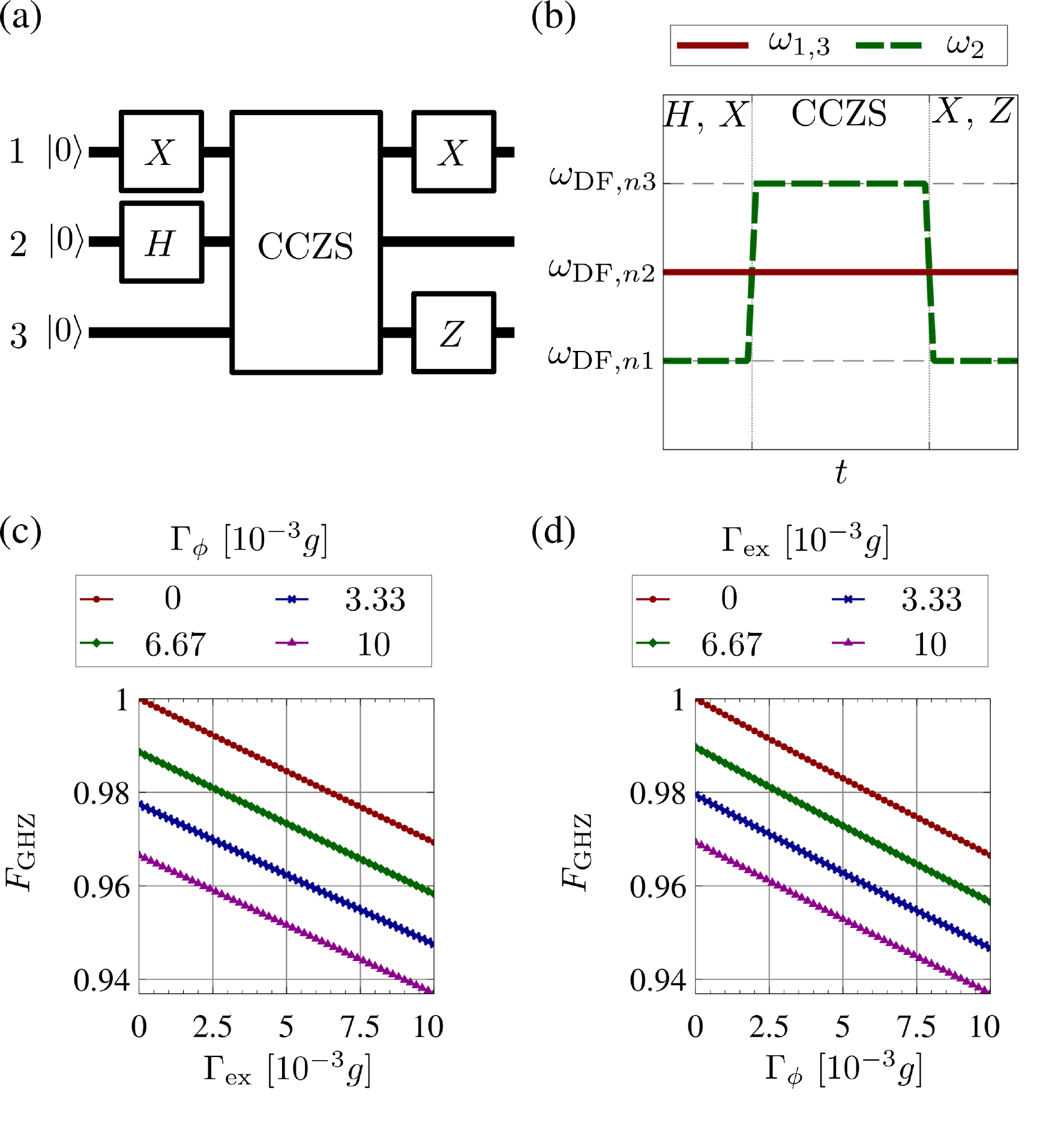}
\caption{Preparation of a three-qubit GHZ state using giant-atom gates. 
(a) Quantum circuit for generating the GHZ state using a CCZS gate and single-qubit gates. 
(b) Frequency-tuning protocol used to activate the required gate interactions of the circuit in (a) via decoherence-free points in the giant-atom setup shown in \figpanel{fig1}{a}.
(c) Fidelity of the resulting GHZ state as a function of qubit decay rate $\Gamma_\text{ex}$ and (d) $\Gamma_\phi$, assuming ideal single-qubit operations.}
\label{fig3}
\end{figure}
 
We first consider the preparation of a three-qubit GHZ state using the setup in \figpanel{fig1}{a}. As shown in \figpanel{fig3}{a}, this task requires only a single CCZS gate and a few single-qubit gates (with two-qubit CZ gates instead of CCZS gates, the same task would require two sequential two-qubit gates, each slower than a three-qubit gate~\cite{Gu2021}). By tuning the frequencies of the giant atoms as illustrated in \figpanel{fig3}{b}, the required interactions can be activated. 

To isolate the contribution of the three-qubit gate, we assume ideal single-qubit operations and focus on the effect of noise during the CCZS gate. This is a reasonable assumption, since single-qubit gates tend to be much faster, and therefore of higher fidelity, than multi-qubit gates. The fidelity of the resulting GHZ state as a function of decay and dephasing rates is shown in \figpanels{fig3}{c}{d}, yielding the approximate expression $F_{\rm GHZ}\approx1-3.17\Gamma_\text{ex}/g-3.35\Gamma_\phi/g$. For representative parameters of $\gamma/ (2\pi) = \qty{4}{\mega\hertz}$, $\Gamma_{\rm ex} = \qty{0.01}{\mega\hertz}$ and $\Gamma_\phi = \qty{0.02}{\mega\hertz}$, the gate time (neglecting single-qubit gates) is approximately \qty{88.4}{\nano\second}, and the resulting state fidelity is \qty{99.61}{\percent}.

\begin{figure}
\center
\includegraphics[width=\linewidth]{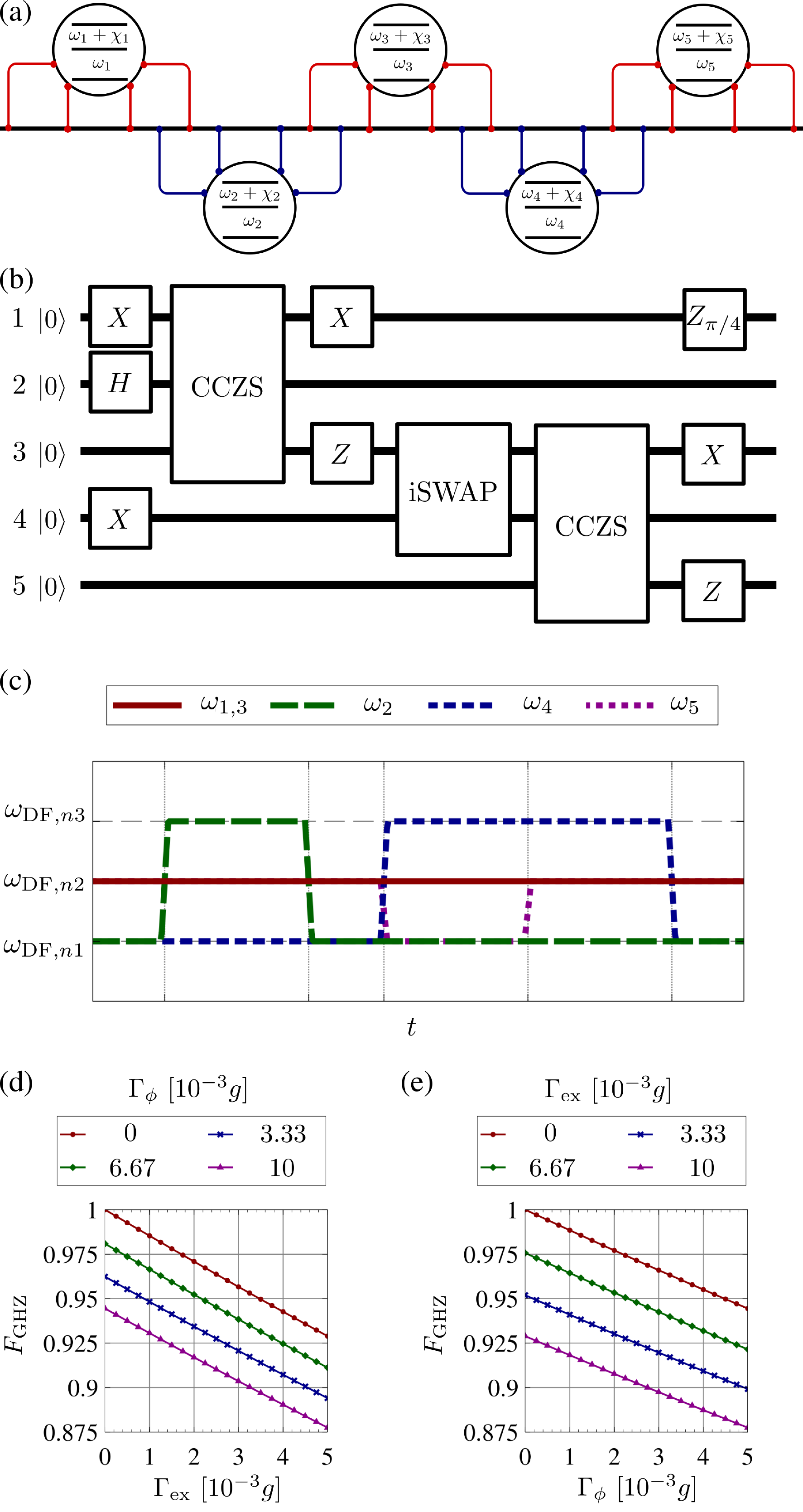}
\caption{Preparation of a five-qubit GHZ state using giant-atom gates. 
(a) Schematic of the setup [an extension of that in \figpanel{fig1}{a}].
(b) Quantum circuit for generating the GHZ state using two CCZS gates, one iSWAP gate, and single-qubit gates. 
(c) Frequency tuning protocol used to activate the required gate interactions of the circuit in (a) via decoherence-free points in the giant-atom setup.
(d) Fidelity of the resulting GHZ state as a function of qubit decay rate $\Gamma_\text{ex}$ and (e) $\Gamma_\phi$, assuming ideal single-qubit operations.}
\label{fig4}
\end{figure}

A key advantage of our architecture is its scalability: additional qubits can be incorporated with minimal overhead by extending the giant-atom chain. This extension enables straightforward preparation of larger GHZ states. As a concrete example, we demonstrate the preparation of a five-qubit GHZ state with the setup in \figpanel{fig4}{a}, using two CCZS gates and one iSWAP gate, in addition to single-qubit operations [see the quantum circuit in \figpanel{fig4}{b}]. The required frequency tuning is shown in \figpanel{fig4}{c}. During the protocol, the frequencies of qubits 2 and 4 need to be shifted by $\omega_0/4$. For a waveguide propagation velocity $v=\qty{1.3e8}{\meter/\second}$~\cite{Goppl2008, Blais2021} and a conservative choice of coupling-point separation $\Delta x=\qty{1}{\cm}$~\cite{Kannan2020, Sundaresan2015}, we have $\omega_0/(2\pi)=\qty{13}{\giga\hertz}$. In current superconducting circuits, qubit frequencies can be tuned at rates as high as $\qty{1}{\giga\hertz} / \qty{}{\nano\second}$~\cite{Collodo2020}. Consequently, the required frequency shifts can be completed within a few nanoseconds, which is negligible compared to the CCZS gate duration. We therefore neglect the contribution of frequency tuning to the total gate time. Again, assuming ideal single-qubit gates, we find that the resulting fidelity scales as [\figpanels{fig4}{d}{e}]: $F_{\rm GHZ, 5}\approx1-14.21\Gamma_\text{ex}/g-11.11\Gamma_\phi/g$. Using the same experimental parameters as above, the total gate time is \qty{239.3}{\nano\second}, and the final state fidelity is \qty{98.54}{\percent}. This performance surpasses that of existing GHZ state preparation experiments~\cite{Warren2023, Pont2024, Zhu2022}, thanks to the reduced circuit depth and the fast, native implementation of entangling gates in our setup.

\section{Conclusion and outlook}\label{sec7}

We have presented a scalable architecture for implementing high-fidelity three-qubit gates using giant atoms coupled to a common waveguide. By leveraging interference effects from spatially separated coupling points, our setup enables native realization of the CCZS and DIV gates through simple frequency tuning, eliminating the need for parametric couplers or complex control schemes. Both gates can be executed on sub-\qty{100}{\nano\second} timescales and achieve fidelities exceeding \qty{99.5}{\percent} using experimentally realistic parameters for superconducting qubits. As an application, we showed how these gates enable fast preparation of three- and five-qubit GHZ states with minimal circuit depth and high fidelity, illustrating the potential of giant-atom platforms for efficient entanglement generation. The intrinsic tunability and low hardware overhead make giant atoms a promising platform for entangled-state preparation and low-depth quantum circuit design in the NISQ era.

Beyond state preparation, the CCZS gate serves as a valuable tool for entanglement routing and multi-qubit interactions in modular quantum architectures~\cite{FuDong2025}. GHZ states, in turn, are widely used as initial states in quantum simulation—for example, to study dynamical phase transitions~\cite{Heyl2018, Zvyagin2016, PhysRevLett.110.135704, Zhang2017, PhysRevApplied.11.044080}, characterize quantum chaos~\cite{PhysRevE.70.016217}, or probe frequency metrology~\cite{Kielinski2024}. The ability of giant atoms to generate such states rapidly and coherently, combined with their compatibility with open-system dynamics~\cite{Chen2025, chen2025scalablequantumsimulatorextended}, positions them as a versatile platform for these quantum-simulation tasks.

Looking ahead, this architecture opens new pathways for implementing native multi-qubit gates in larger systems, enabling the scalable preparation of complex entangled states and advancing the scope of dynamical quantum simulation. Further experimental realizations, likely using superconducting circuits, could pave the way toward modular quantum processors with reduced circuit depth and enhanced noise resilience. In this context, optimal control techniques---including pulse shaping, gradient-based optimization, and robust gate design---can help further improve fidelities, suppress crosstalk, and mitigate leakage. These methods will be especially valuable in scaling up to larger systems where coherence and control demands become more stringent. The combination of native three-qubit gates, flexible connectivity, and optimal control in giant atom platforms may ultimately unlock new regimes in quantum simulation that go beyond what is accessible with conventional architectures.


\begin{acknowledgements}
    
The numerical calculations were performed using the QuTiP library~\cite{Johansson2012, Johansson2013, Lambert2024}; source codes are available on GitHub~\cite{SourceCode}. G.C.~is supported by European Union's Horizon Europe programme HORIZON-MSCA-2023-PF-01-01 via the project 101146565 (SING-ATOM). A.F.K.~acknowledges support from the Swedish Foundation for Strategic Research (grant numbers FFL21-0279 and FUS21-0063), the Horizon Europe programme HORIZON-CL4-2022-QUANTUM-01-SGA via the project 101113946 OpenSuperQPlus100, and from the Knut and Alice Wallenberg Foundation through the Wallenberg Centre for Quantum Technology (WACQT).

\end{acknowledgements}


\bibliography{main,GA,NH_bib}

\end{document}